\documentclass[aps,twocolumn,prb,showpacs]{revtex4}
\usepackage{graphicx}
\usepackage{dcolumn}
\usepackage{amssymb}
\usepackage{times}
\usepackage{amsmath}
\usepackage{amsfonts}
\usepackage{mathrsfs}
\usepackage{setspace}
\usepackage{epsfig}
\usepackage{latexsym}
\usepackage{bbm}
\usepackage{subfigure}
\usepackage{float}
\usepackage{flafter}
\usepackage{bm}
\usepackage{epstopdf}
\usepackage{hyperref}
\usepackage{color}

\begin{document}

\title{Quasiparticle collapsing in an anisotropic $t$-$J$ ladder}
\author{Zheng Zhu$^{1}$ and Zheng-Yu Weng$^{1,2}$}
\affiliation{$^{1}$ Institute for Advanced Study, Tsinghua University, Beijing, 100084, China}
\affiliation{$^{2}$ Collaborative Innovation Center of Quantum Matter, Tsinghua University, Beijing, 100084, China}

\date{\today}



\begin{abstract}
Quasiparticle collapsing is a central issue in the study of strongly correlated electron systems. In the one-dimensional case, the quasiparticle collapsing in a form of spin-charge separation has been well established, but the problem remains elusive in dimensions higher than one. By using density matrix renormalization group (DMRG) algorithm, we show that in an anisotropic two-leg $t$-$J$ ladder, an injected single hole behaves like a well-defined quasiparticle in the strong rung limit, but undergoes a ``phase transition'' with the effective mass diverging at a quantum critical point (QCP) towards the isotropic limit. After the transition, the quasiparticle collapses into a loosely bound object of a charge (holon) and a spin-1/2 (spinon), accompanied by an unscreened phase string as well as a substantially enhanced binding energy between two doped holes. A phase diagram of multi-leg ladders is further obtained, which extrapolates the QCP towards the two-dimensional limit.
The underlying novel mechanism generic for any dimensions is also discussed.
\end{abstract}

\maketitle

The Landau's Fermi liquid theory is characterized by the low-lying quasiparticle excitation that carries well-defined momentum, charge, spin, and a renormalized effective mass. The collapse of such a quasiparticle excitation will be a hallmark of a non-Fermi-liquid state. In particular, the breakdown of the quasiparticle in a form of spin-charge separation has been conjectured in the study of doped Mott insulators, notably the high-$T_c$ cuprates\cite{Anderson87,Kivelson87,Zou88,Nagaosa98,Senthil00,Lee06,Weng1997,Weng2011}. However, no consensus has been reached yet on how a quasiparticle precisely falls into parts in such strongly correlated electron systems.

A $t$-$J$ square ladder as a quasi one-dimensional (1D) doped Mott insulator system has been intensively investigated\cite{Dagotto92,Dagotto96,Hayward96, Troyer96,Sigrist94,White94,White97,Oitmaa99,Weng99,Sorella02,Becca02,ZZ2013,ZZ2014}. Such systems are beyond a purely 1D system due to the presence of closed loops of various sizes, and can be accurately studied by the DMRG numerical method\cite{DMRG92}. Experimentally, there are also several available materials with the ladder structure \cite{2legMaterial}. Because of the peculiar quantum destructive interference in the closed paths,  a DMRG study has recently revealed\cite{ZZ2013} a generic self-localization of a single hole injected into the spin ladders in the isotropic limit. It implies the failure of a conventional quasiparticle picture in a way very distinct from a purely 1D system\cite{Luttinger}.

In this Letter, we focus on a two-leg $t$-$J$ ladder system in which the undoped spin background remains gapped. By using DMRG, we find that for an injected hole, the quasiparticle description is restored if the ladder is in an anisotropic (strong rung) regime. Then, as the ladder anisotropic parameter is continuously tuned from strong rung coupling towards the isotropic limit, there exists a QCP, at which the quasiparticle collapses with its effective mass diverges. Subsequently the doped hole fractionalizes into a composite structure as a bound state of an incoherent holon and a deconfined spinon. The momentum distribution of the hole also exhibits a qualitative change across the QCP. The underlying microscopic mechanism responsible for the fractionalization of the hole will be discussed. Interestingly the binding energy of two holes also gets substantially enhanced after the quasiparticle collapsing. Such a QCP is further shown to persist with the increase of the leg-number of the ladders,
which may shed light to the understanding of the quasiparticle collapsing and pairing in the two-dimensional (2D) doped Mott insulator.
\begin{figure}[tbp]
\centerline{
    \includegraphics[height=1in,width=3.2in] {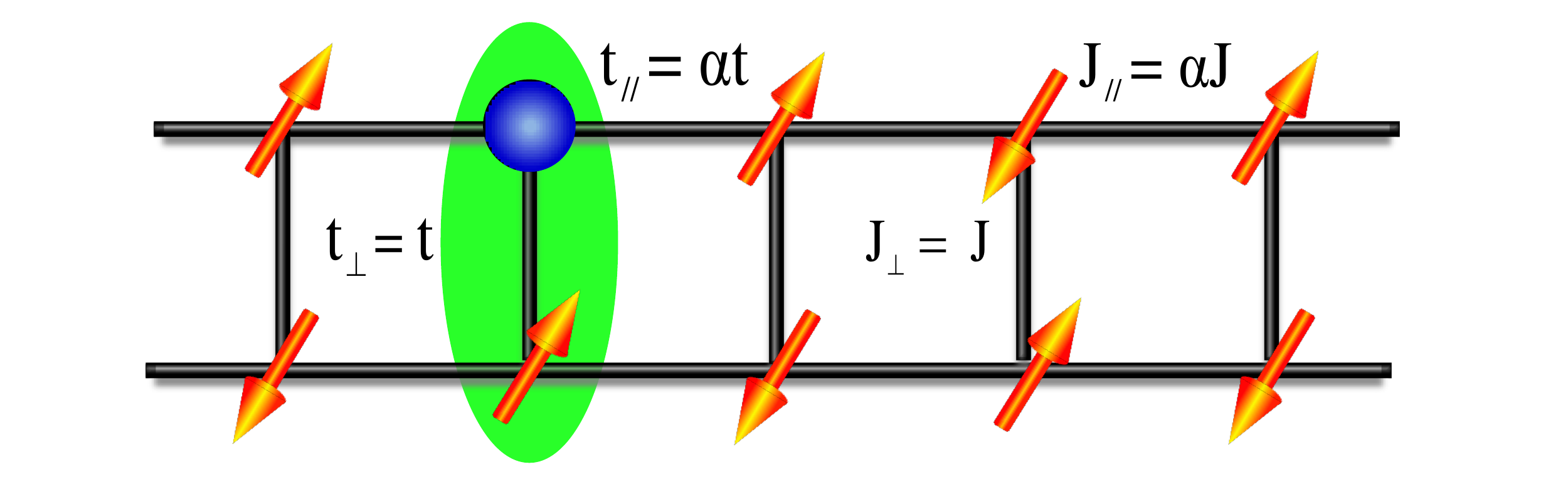}
    }
\caption{(Color online) The parameters of the anisotropic $t$-$J$ model on a two-leg square ladder. Here,  $t_ {\bot}= t$ ($t_{\parallel}=\alpha t$) and $J_ {\bot}= J$ ($J_{\parallel}=\alpha J$) describe the inter-chain (intra-chain) hopping and superexchange couplings, respectively. At $\alpha=1$, it reduces to the isotropic limit.} \label{Fig:Model}
\end{figure}

The $t$-$J$ Hamiltonian $H=H_{t}+H_{J}$ for an anisotropic two-leg ladder system is composed of four terms: $H_{t_ {\bot}}+H_{t_{\parallel}}+H_{J_ {\bot}}+H_{J_ {\parallel}}$ given by
\begin{equation}
\begin{split}
H_{t_ {\bot}}&=  - t_ {\bot} \sum_{i,y=0,\sigma } {(c_{i,y,\sigma }^\dag {c_{i,y+1,\sigma }} + h.c.)},\\
H_{t_{\parallel}}&= - t_{\parallel}\sum_{i,y,\sigma } {(c_{i,y,\sigma }^\dag {c_{i + 1,y,\sigma }} + h.c.)},\\
H_{J_ {\bot}} &=  J_ {\bot} \sum_{i,y=0} {(\mathbf{S}_{i,y}\cdot \mathbf{S}_{i,y+1}-\frac{1}{4}n_{i,y}n_{i,y+1})} ,\\
H_{J_ {\parallel}} &= J_{\parallel}\sum_{i,y} {(\mathbf{S}_{i,y}\cdot \mathbf{S}_{i+1,y}-\frac{1}{4}n_{i,y}n_{i+1,y})}.
\end{split}
\label{Eq:SquaretJModel}
\end{equation}
on a two-leg ladder with the total site number $N=N_{x}\times N_{y}$ ($N_{y}=2$) as sketched in Fig.~\ref{Fig:Model}. In Eq. (\ref{Eq:SquaretJModel}), the summation over $i$ along the chain direction runs over all rungs, $y$ ($=0,1$) and $\sigma$ are leg and spin indices, respectively.  ${c_{i,y,\sigma }^{\dagger }}$ is the electron creation operator and ${\mathbf{S}_{i,y}}$ the spin operator at site ($i$,$y$). The Hilbert space is always constrained by the no-double-occupancy condition, i.e., the number operator $n_{i}\leq 1$. Here, $H_{t_ {\bot}}$ ($H_{t_{\parallel}}$) and $H_{J_ {\bot}}$ ($H_{J_ {\parallel}}$) describe the inter-chain (intra-chain) hole hopping and spin superexchange interaction, respectively. For simplicity, in the following we shall fix $t_{\bot}/J_{\bot}=t_{\parallel}/J_ {\parallel}=3$, or equivalently, take  $t_{\bot}\equiv t$, $J_{\bot} \equiv J$, $t_{\parallel}\equiv \alpha t$, $ J_ {\parallel}\equiv \alpha J$ with $t/J=3$.  For the present simulation, we use $U(1)$ invariant code and set $J$ as the unit of energy. We keep up to 4000 states in each DMRG block with around 10 to 50 sweeps, and this is proved to be enough to give excellent convergence with the truncation error is of the order or less than $10^{-8}$ .  Then we continuously tune  $\alpha$ from $0$ to $1$ between the strong rung and isotropic limits as illustrated in Fig.~\ref{Fig:Model}. At half-filling, the system remains spin-gapped without a phase transition, and in particular, the ground state simply reduces to a direct product of spin-singlet rungs in the strong rung limit of $\alpha\rightarrow 0$.\cite{Sachdev90}

\begin{figure}[tbp]
\centerline{
    \includegraphics[height=2in,width=2.8in] {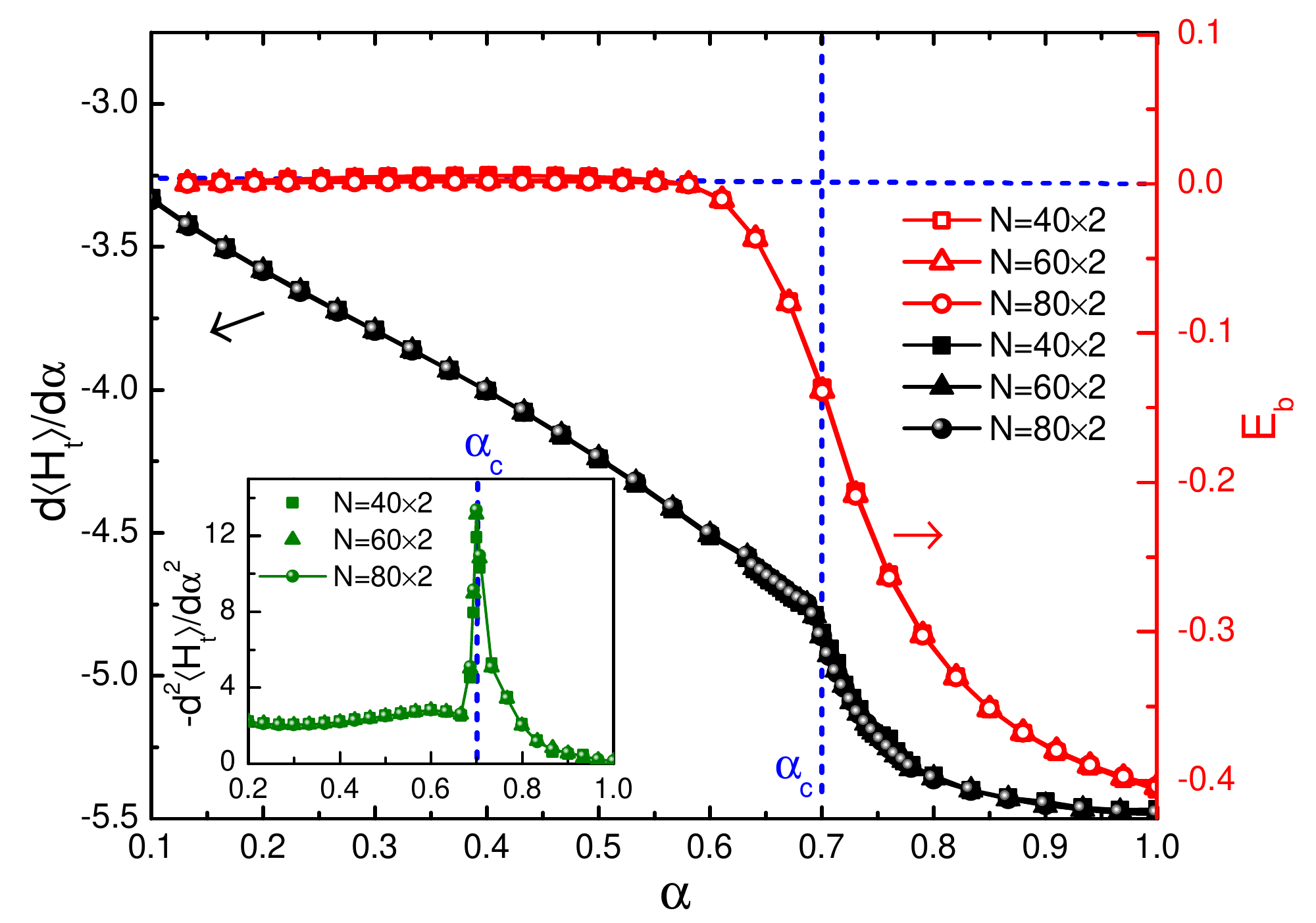}
    }
\caption{(Color online)  The first-order derivative of the kinetic energy $\langle H_{t}\rangle$ and the second-order derivative (the inset) indicate the presence of a quantum critical point at $\alpha (\equiv t_ {\parallel}/t_ {\bot})=\alpha_c \sim 0.7$ (with $t_{\bot}/J_ {\bot}=t/J=3$).  The two-hole pairing is also substantially enhanced at $\alpha>\alpha_c$ as shown by the binding energy $E_b$ .  } \label{Fig:TJtransition}
\end{figure}

Now consider the one-hole-doped case. As shown in Fig. ~\ref{Fig:TJtransition} and the inset,  a QCP is clearly indicated at $\alpha=\alpha_c\sim 0.7$  by the first- and second-order derivatives of the kinetic energy $\langle H_{t} \rangle$ over $\alpha $. (Note that the derivatives of the superexchange energy $\langle H_{J} \rangle$ remain smooth without a singularity, which is not shown in the figure.) What we shall establish first below, is that  at  $\alpha< \alpha_c  $, the single doped hole behaves like a Bloch quasiparticle, which possesses a well-defined momentum, effective mass, charge, spin, and finite quasiparticle weight. In fact, at strong rung limit $\alpha \ll 1$, the quasiparticle behavior can be well described by a perturbation theory\cite{Kivelson14}. But beyond the critical point $\alpha_c $, the quasiparticle picture of the single doped hole will break down completely.

By contrast,  when two holes are injected into the gapped two-leg spin ladder, they always form a binding state in the quasiparticle collapsing regime. The pairing even persists into the quasiparticle regime with reducing binding strength, which eventually vanishes around $\alpha \sim 0.6$ as shown by the binding energy $E_{b}$ in Fig. ~\ref{Fig:TJtransition} (red circles). Here the binding energy is defined by $E_{b}\equiv E^{\text{2-hole}}_{G}+E^{0}_G-2E^{\text{1-hole}}_{G}$, where $E^{\text{2-hole}}_{G}$, $E^{\text{1-hole}}_G$, and $E^0_G$ denote the ground-state energies of the two-hole, one-hole, and undoped states, respectively.

\begin{figure}[tbp]
\centerline{
    \includegraphics[height=5.6in,width=2.8in] {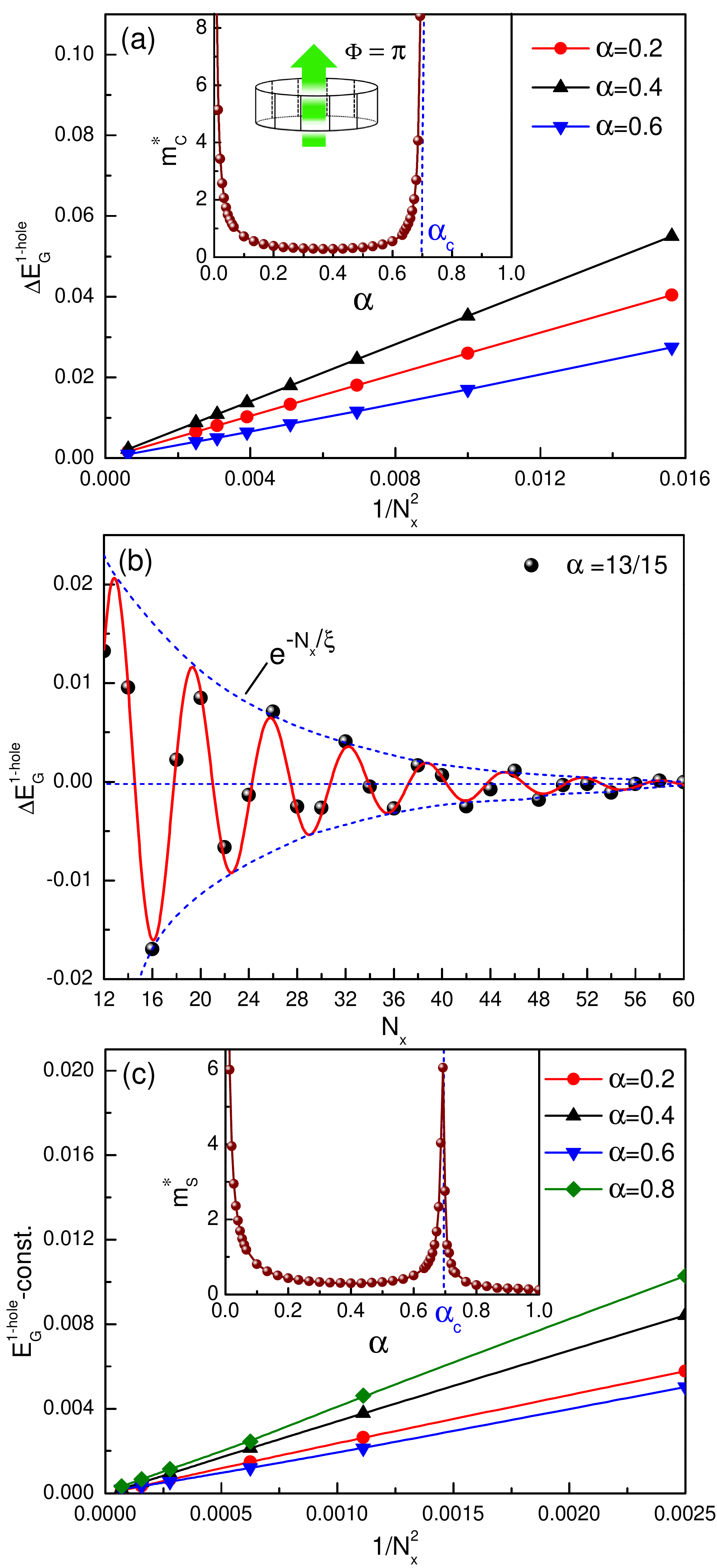}
    }
\caption{(Color online) Effective mass of the quasiparticle is well-defined at $\alpha<\alpha_c$, but is divergent at the quantum critical point $\alpha_c$. (a) $\Delta E_{G}^{\text{1-hole}}$ [defined in Eq. (\ref{dE})] exhibits a \textquotedblleft free particle\textquotedblright \ behavior:  $\Delta E_{G}^{\text{1-hole}}  \propto  1/N_x^2$, in a loop of the circumference  $N_x$ (cf. the inset). The effective mass $m^{\ast} \propto m_c^{\ast}$ with $1/m_c^{\ast}$ defined as the slope of $1/N_x^2$ [shown in the inset of (a), in which $m_c^{\ast}$ diverges at $\alpha_c$]; (b) At $\alpha>\alpha_c$, $\Delta E_{G}^{\text{1-hole}}$ oscillates and decays exponentially with $m_c^{\ast}=\infty$ [presented in (b) is the case at $\alpha=13/15$ with the charge localization length\cite{ZZ2013} $\xi\sim 12.6$]; (c) The one-hole ground state energy $E_{G}^{\text{1-hole}}$ calculated under an open boundary of length $N_x$. Here the slope of $E_{G}^{\text{1-hole}}$ (subtracted by a constant term) defines another effective mass $m_{s}^{\ast}$ shown in the inset of (c), which is essentially the same as $m_c^{\ast}$ at $\alpha<\alpha_c$. But $m_c^{\ast}$ and $m_s^{\ast}$ differ completely at $\alpha>\alpha_c$, suggesting the charge-spin separation (see the text). } \label{Fig:1hole_Edifference}
\end{figure}

\begin{figure}[tbp]
\centerline{
    \includegraphics[height=5.5in,width=2.8in] {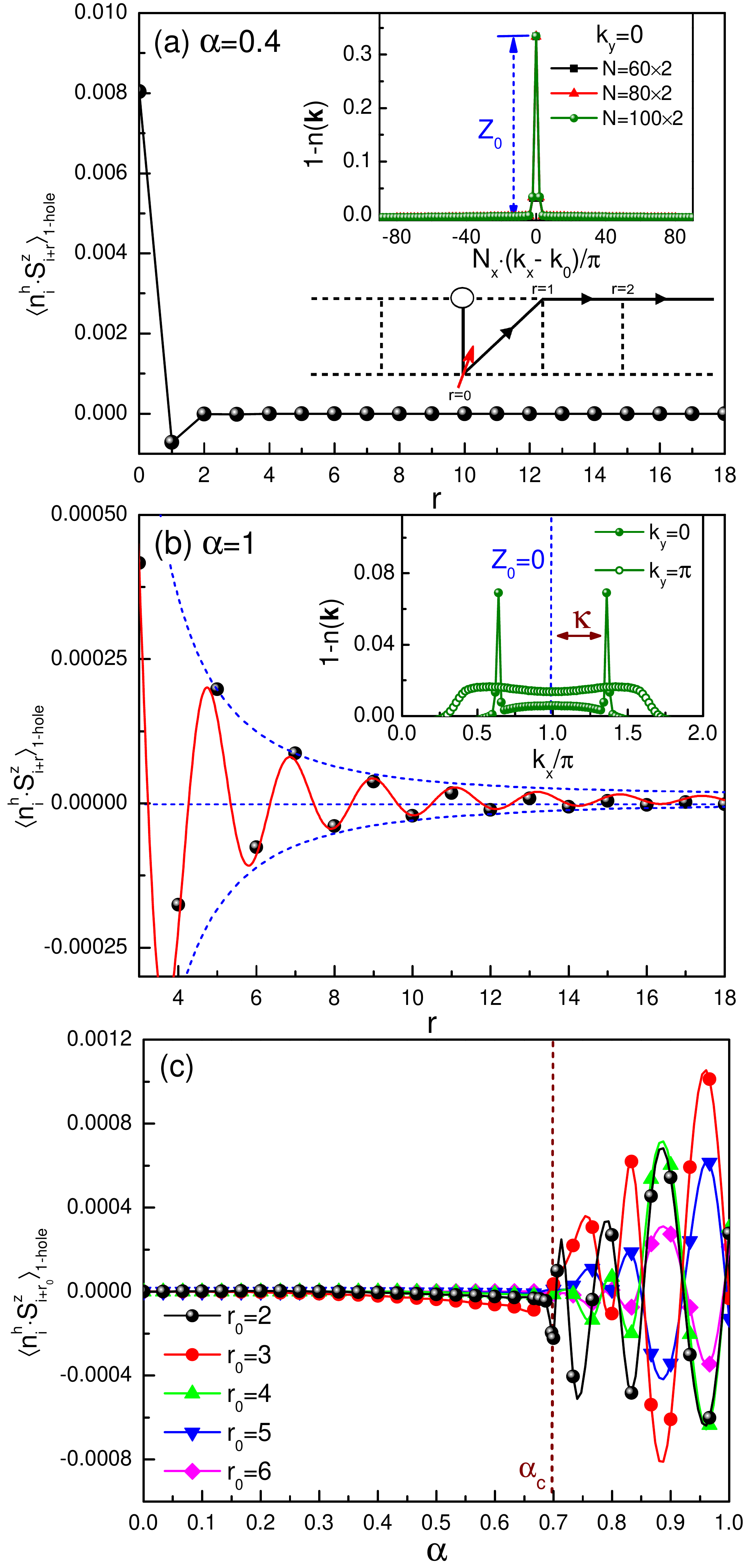}
    }
\caption{(Color online) Spin-charge correlator $\langle n_{i}^h \cdot S_{i+r}^z\rangle$ with the labeling $r$ defined in (a). The tight-binding of the holon-spinon inside the quasiparticle is shown in (a) for $\alpha=0.4<\alpha_c$ with $N=60\times2$. The inset of (a): the momentum distribution of the hole satisfies a scaling law: $k_x \rightarrow N_x (k_x-k_0)$, indicating a well-defined momentum at $k_0=\pi$ with a finite spectral weight $Z_{0}$\cite{ZZ2013}; (b) The fractionalization of the quasiparticle occurs at $\alpha>\alpha_c$ with a composite structure of loosely bound charge and spin as shown by the spin-charge correlator ($N=60\times2$). The inset of (b): the momentum distribution is fundamentally changed with $Z_0$ vanishing at $k_0$ ($N=100\times2$) ; (c) A sharp increase of the amplitude for the hole-spin separation at $r \geq 2$ as $\alpha\geq \alpha_c$. } \label{Fig:spin-charge_cor}
\end{figure}

For the single hole case, a finite effective mass at $\alpha< \alpha_c$ is identified in the inset of  Fig.~\ref  {Fig:1hole_Edifference} (a). Here, to determine the effective mass of the charge, the two-leg ladder is made of a loop along the long chain direction with a magnetic flux $\Phi$ threading through [cf. the inset of of Fig.~\ref  {Fig:1hole_Edifference} (a)]. Then the ground state energy difference between $\Phi=\pi$ and $0$, i.e.,
\begin{equation}\label{dE}
 \Delta E_{G}^{\text{1-hole}}\equiv E_{G}^{\text{1-hole}}(\Phi=\pi )-E_{G}^{\text{1-hole}}(\Phi=0),
 \end{equation}
corresponds to the energy difference under the change of the boundary condition from the periodic to anti-periodic one for the charge (hole). If the doped hole behaves like a \textquotedblleft Bloch quasiparticle\textquotedblright , $\Delta E_{G}^{\text{1-hole}}$ is expected to be proportional to $1/N_{x}^2$, with the inverse of the slope $m_c^{\ast}$ proportional to the effective mass.

As shown in Fig.~\ref  {Fig:1hole_Edifference} (a), a finite $m_c^{\ast}$ is indeed obtained at $0<\alpha <\alpha_c$ (which diverges at  $\alpha =0$ because of the vanishing inter-rung hopping). Then $m_c^{\ast}$ diverges again approaching the critical point $\alpha_c$  [cf. the inset of Fig.~\ref  {Fig:1hole_Edifference} (a)]. Beyond $\alpha_c$, $\Delta E_{G}^{\text{1-hole}}$ starts to oscillate and decay exponentially as a function of $N_{x}$ as illustrated in Fig.~\ref  {Fig:1hole_Edifference} (b), with the disappearance of the term proportional to $1/N_{x}^2$. It implies the self-localization of the doped hole \cite{ZZ2013} with the effective mass $m_c^{\ast}=\infty$ at $\alpha \geq\alpha_c$.

On the other hand, the effective mass can be also determined alternatively. Fig.~\ref  {Fig:1hole_Edifference} (c) shows the one-hole ground state energy $E_{G}^{\text{1-hole}}$ calculated under the fully open boundary condition. Besides a constant term, $E_{G}^{\text{1-hole}}$ can be also well fitted by ${m_{s}^{\ast}}^{-1}/N_{x}^{2}$, with $m_{s}^{\ast}$ essentially the same as $m_c^{\ast}$ at $\alpha<\alpha_c$ as shown in the inset of Fig.~\ref  {Fig:1hole_Edifference} (c). One finds that $m_{s}^{\ast}$ also diverges at $\alpha_c$. However, in contrast to $m_c^{\ast}$, $m_{s}^{\ast}$ becomes finite again at $\alpha>\alpha_c$. Namely, in opposite to the charge part of the doped hole (holon) being localized at $\alpha>\alpha_c$, a charge-neutral gapless excitation (spinon) is still present in this regime.
\begin{figure*}[tbp]
\centerline{
    \includegraphics[height=4in,width=6in] {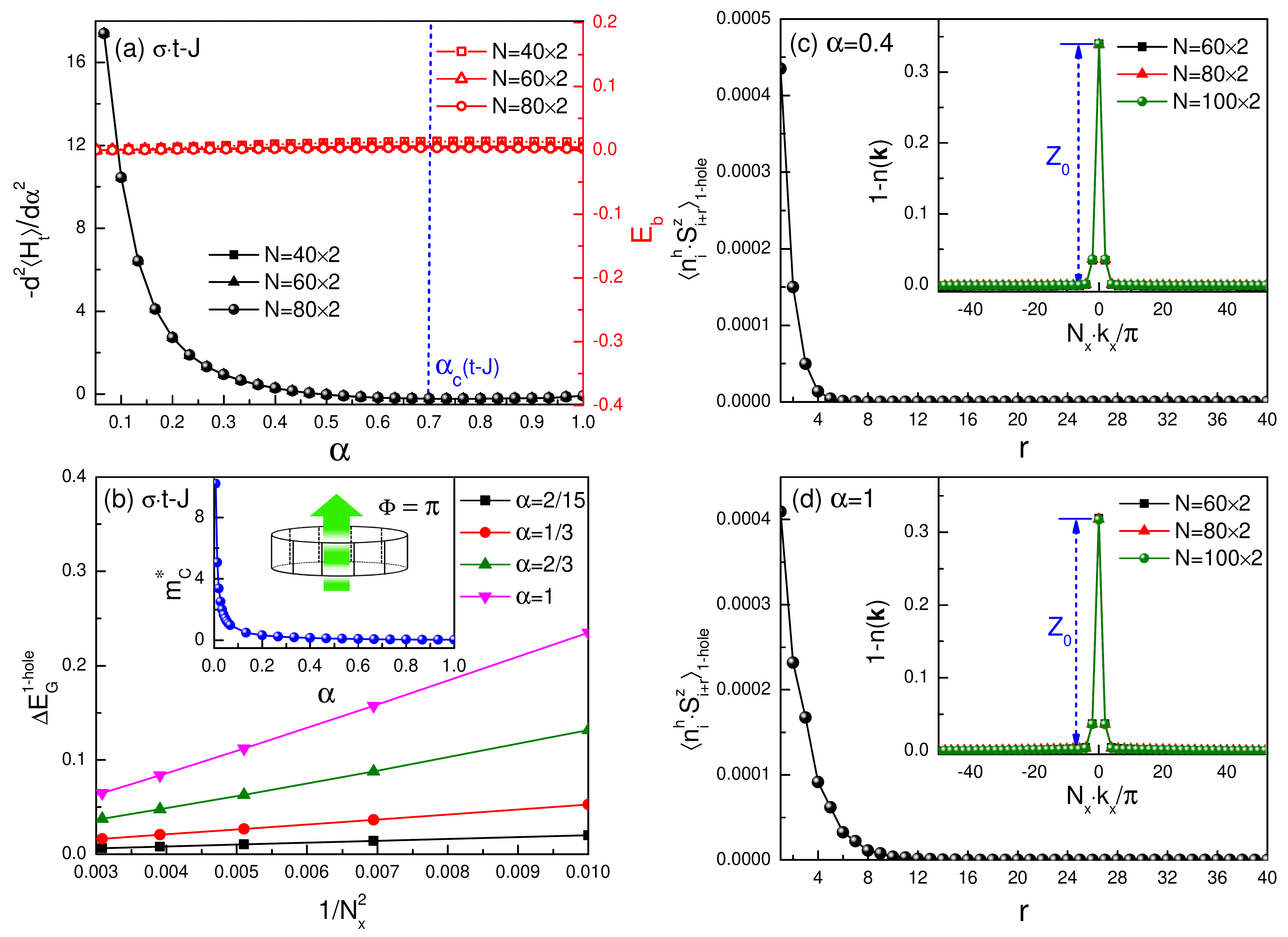}
    }
\caption{(Color online) The quantum critical point $\alpha_c$ disappears in the $\sigma$$\cdot t$-$J$ model (see the text) with the quasiparticle description valid throughout the whole regime of $\alpha$. (a) The second-order derivative of the single hole kinetic energy does not show any singularity at a finite $\alpha$. Correspondingly, the two-hole binding energy remains weak similar to the quasiparticle regime in Fig. ~\ref{Fig:TJtransition} ; (b) The effective mass determined by the slope of $1/N_{x}^2$ behavior of $\Delta E_{G}^{\text{1-hole}}$ remains smooth and finite at $\alpha>0$ (the inset); (c) and (d): the integrity of the quasiparticle is ensured by the spin-charge binding at different $\alpha$'s ($N=80\times2$). The insets of (c) and (d):  the momentum distribution for the single hole is similar to the one in the quasiparticle regime of the $t$-$J$ model in Fig.~\ref {Fig:spin-charge_cor} (a). The scaling law ensures a well-defined momentum at $k_x=0$. } \label{Fig:SigmaTJ}
\end{figure*}

The sharp contrast between $m_c^{\ast}$ and $m_{s}^{\ast}$ suggests that the quasiparticle collapses at $\alpha>\alpha_c$ by a specific form of the electron fractionalization. One may directly measure the spin-charge separation by calculating the spin-charge correlator $\langle n_{i}^h \cdot S_{i+r}^z\rangle$. As shown in Fig.~\ref {Fig:spin-charge_cor} (a) ($\alpha=0.4<\alpha_c$),  the spin and charge are tightly bound together at a length scale of one lattice constant. Such a stable hole object has a well-defined mass $m^{\ast}$ and behaves like a Bloch wave with a definite momentum. The momentum distribution 1-$n(\mathbf{k})$ of the hole is presented in the inset of Fig.~\ref {Fig:spin-charge_cor} (a). Here
$n(\mathbf{k})\equiv \sum_{\sigma }\langle {c_{\mathbf{k}\sigma }^{\dag }c_{\mathbf{k}\sigma }}\rangle$, which can be obtained by a Fourier transformation of $\sum_{\sigma }\langle {c_{i\sigma }^{\dag}c_{j\sigma }}\rangle$. The inset of Fig.\ref{Fig:spin-charge_cor} (a) shows that the hole momentum distribution as a universal curve after the rescaling  $k_x-k_0 \rightarrow N_x \cdot (k_x-k_0)$ with $k_0=\pi$, indicating that the hole in the ground state possesses a well-defined momentum $(k_0, k_y=0)$ with a finite quasiparticle spectral weight $Z_{0}$ in the thermodynamic limit.

The quasiparticle collapsing at $\alpha>\alpha_c$ is in a form of fractionalization as shown in Fig.~\ref {Fig:spin-charge_cor} (b)  (at $\alpha=1$), where the spin-charge correlator oscillates and decays much slower. Corresponding,  the quasiparticle weight $Z_{0}=0$ at $k_0=\pi$ and the hole momentum distribution is qualitatively changed as presented in the inset of  Fig.~\ref {Fig:spin-charge_cor} (b) with two new peaks emerging at $k_x=k_0\pm \kappa$ and $k_y=0$ with $\kappa$ depending on $\alpha$ and $t/J$.

Figure ~\ref {Fig:spin-charge_cor} (c) further illustrates how the quasiparticle fractionalizes. At  $\alpha< \alpha_c$, the amplitude for  the spin-charge separation distance $r\geq 2$ is exponentially small, implying the tight-binding of the holon-spinon within the quasiparticle at $r< 2$ in Fig. ~\ref {Fig:spin-charge_cor} (a). But at $\alpha\geq \alpha_c$ a sharp arise of the amplitude at $r\geq 2$ indicate the emergence of a composite structure for the quasiparticle as the spin partner can now move away from the holon over a larger distance as shown in Fig.~\ref {Fig:spin-charge_cor} (b).

To understand the underlying physics of the quasiparticle collapsing, we slightly modify the hopping terms $H_{t_ {\bot}}$ and $H_{t_{\parallel}}$ in Eq. (\ref{Eq:SquaretJModel}) by introducing a sign prefactor $\sigma =\pm$ such that  $c_{i\sigma}^{\dag }c_{j\sigma }\rightarrow \sigma
c_{i\sigma}^{\dag }c_{j\sigma }$. This is a generalization of the so-called $\sigma$$\cdot$$t$-$J$ model in the isotropic limit, where the hopping term $H_t$ is replaced by\cite{ZZ2013}: $H_{\sigma \cdot t} = -t \sum_{\langle {ij}\rangle \sigma }\sigma
{({c_{i\sigma}^{\dag }c_{j\sigma }+h.c.})}$.
Then we can carry out the same DMRG calculation, and as clearly indicated in Fig.~\ref{Fig:SigmaTJ} (a), the QCP $\alpha_c$ simply disappears. Namely, there is no more quasiparticle collapsing and there exists only one phase continuously interpolating between the isotropic and strong rung limits. Figure ~\ref{Fig:SigmaTJ} illustrates that the single hole moving in the gapped spin background always keeps its quasiparticle identity with a well-defined momentum at $k_x=0$ (note that it is different from $k_0$ in the $t$-$J$ ladder case) with a finite spectral weight, a finite effective mass $m_c^{\ast}$, and the spin-charge confinement. As one can see from Fig.~\ref{Fig:SigmaTJ} (c), even in the isotropic limit of $\alpha=1$, the hole still keeps the integrity of a Bloch quasiparticle with charge and spin tightly bound. As a matter of fact, we have checked that the same phase still persists at $\alpha\gg1$. Furthermore, the binding energy is also substantially weakened in the whole regime [cf. Fig.~\ref{Fig:SigmaTJ} (a)], similar to the quasiparticle regime in the $t$-$J$ ladder case.

Previously it has been demonstrated\cite{ZZ2013} that the sole distinction between the isotropic $t$-$J$ and $\sigma$$\cdot$$t$-$J$ models lies in the so-called phase string \cite{Sheng1996,Wu-Weng-Zaanen} associated with each path of the hole motion, which is present in the former but is precisely removed in the latter. The same proof remains true in the present anisotropic ladder case.
Such a phase string represents a singular phase shift produced by the scattering between the spin background and doped charge
\cite{Sheng1996,Weng1997,Wu-Weng-Zaanen,Zaanen_09} for general dimensions of bipartite lattice. Its destructive quantum interference has been previously found to
lead to the localization of the doped hole in the isotropic limit $\alpha=1$ of the $t$-$J$ ladder with the leg-number $N_y>1$ \cite{ZZ2013}.
The phase string is also responsible for the strong binding found in the quasiparticle collapsing regime of the $t$-$J$ model (cf. Fig. ~\ref{Fig:TJtransition}), as has been carefully examined in the isotropic case\cite{ZZ2014} before.

\begin{figure}[tbp]
\centerline{
    \includegraphics[height=2in,width=3in] {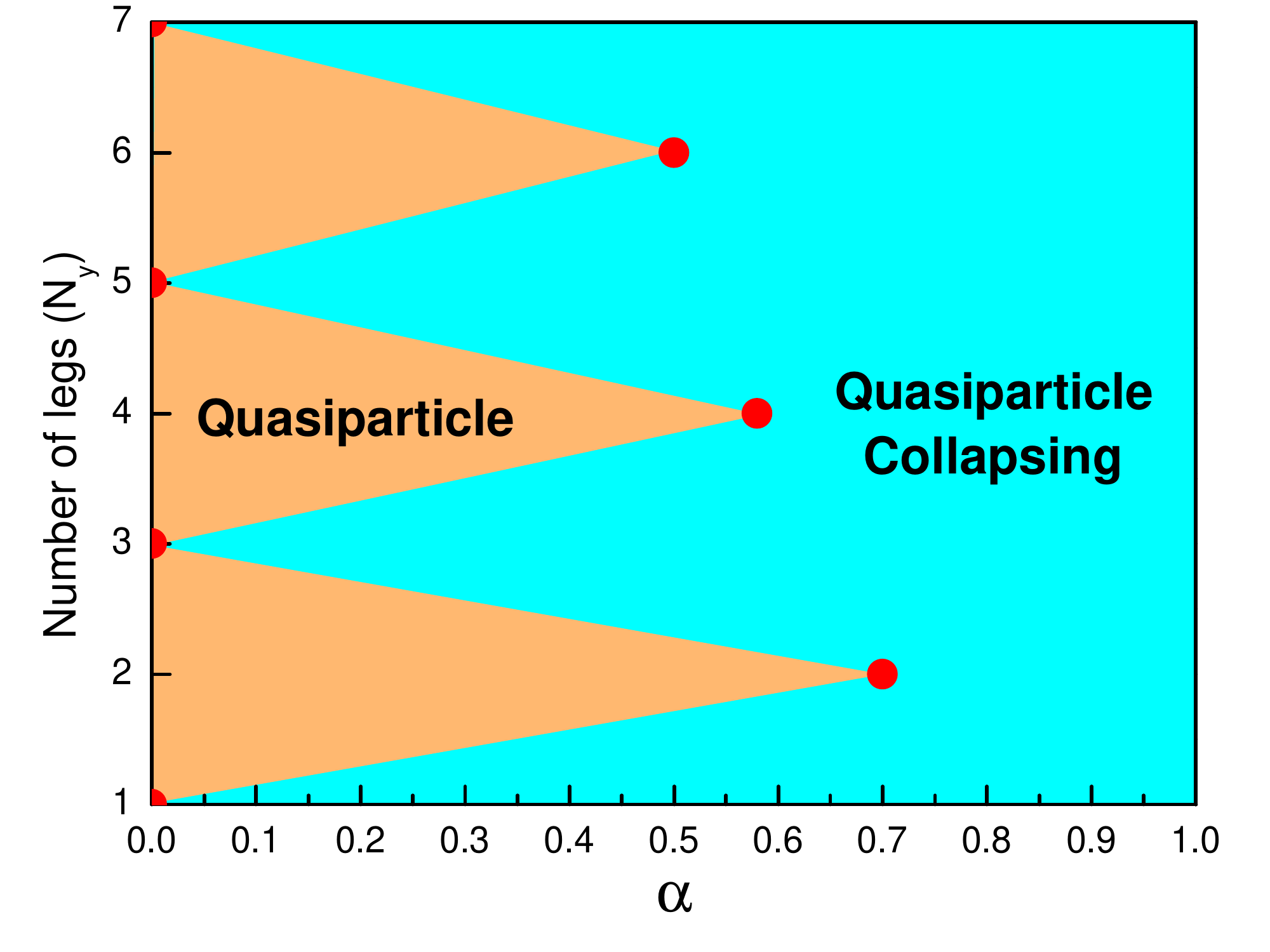}
    }
\caption{(Color online) The QCP (red dot) separating the quasiparticle (orange-colored) and quasiparticle collapsing (blue-colored) regimes is shown as a function of the anisotropy parameter $\alpha$, for the $t$-$J$ ladders of the leg number $N_y$ from $1$ to $7$ (with $t/J=3$).  } \label{Fig:Phasediagram}
\end{figure}

Finally, the QCP $\alpha_c$ is systematically calculated for the multi-leg $t$-$J$ ladders as shown in Fig. ~\ref{Fig:Phasediagram}. Here $\alpha_c$ is determined by the singularity in the ground state energy similar to that for the two-leg ladder shown in the inset of Fig. ~\ref{Fig:TJtransition}. Similar to the two-leg case, in the multi-leg ladders the horizontal chains are glued by a fixed hopping integral $t$ and a superexchange coupling $J$ perpendicular to the chain direction. Physically,  $\alpha_c$ separates the non-degenerate quasiparticle state from a quasiparticle collapsing state. For an odd-leg spin ladder, the spin background always remains gapless at half-filling and generally $\alpha_c=0^+$ is found in the single-hole state where a true spin-charge separation persists. By contrast, as our above study of the two-leg ladder has clearly shown, in the presence of a spin gap in an even-leg ladder, the singular phase string effect may get ``screened'' via a tight-binding of the charge and spin partners to form a coherent Bloch-type quasiparticle, at least in the strong rung limit of $\alpha\ll 1$. With the reducing spin gap by increasing $\alpha$ or leg-number $N_y$, the tight binding between the holon and its backflow spinon gets weakened, eventually resulting in quasiparticle collapsing at some $\alpha_c$, where the holon and spinon form a {\em loosely bound state} (instead of a simple spin-charge separation in the odd-leg cases) with an unscreened and irreparable phase string reemerging to accompany the motion of hole. In fact, a finite $\alpha_c$ does persist in all the even-leg ladders shown in Fig. ~\ref{Fig:Phasediagram}, which monotonically decreases with the increase of the leg numbers up to $N_y=6$. A microscopic wave function approach to this problem will be presented elsewhere.

\emph{Note added}. After the submission of the present paper, we became aware of a DMRG study of the same two-leg $t$-$J$ ladder doped by one hole\cite{White15}, in which the authors have confirmed the existence of $\alpha_c\sim 0.7$, the divergence of the effective mass $m^*_s$ at  $\alpha_c$, the incommensurate momentum split together with the enlarged spin-charge separation at $\alpha>\alpha_c$ found in this work. However, we notice that the physical interpretation of the nature at $\alpha>\alpha_c$ in that paper is different from the current picture of Bloch quasiparticle collapsing. We point out that our interpretation is further supported based on some additional DMRG probes including the charge response to the inserting flux, the $\sigma$$\cdot$$t$-$J$ model without the phase string effect, etc., which are absent in that work.

\textbf{Acknowledgement} Stimulating and useful discussions with G. Baskaran, L. Fu, S. Kivelson, D.H. Lee, P.A. Lee, S.S. Lee, N. Nagaosa, T. L. Ho, D. N. Sheng, X. G. Wen, J. Zaanen, and especially H.-C. Jiang are acknowledged. This work was supported by  the NBRPC Grant no. 2010CB923003.


\end{document}